\documentclass[twocolumn, a4paper, notitlepage]{article}
\usepackage{natbib}
\usepackage{graphicx}
\usepackage[colorlinks=true]{hyperref}
\providecommand{\grl}{Geophys. Res. Lett.}
\providecommand{\altaffilmark}[1]{\textsuperscript{#1}}

\pdfoutput=1
\setcitestyle{square}

\makeatletter
\renewcommand{\maketitle}{
    \begin{center}
      \large
      {\LARGE\@title}
      \par\vspace{1ex}
        \@author
	\par\vspace{1ex}
      \@date
    \end{center}
    \@thanks
}
\makeatother

\title{Plasma observations during the Mars atmospheric ``plume'' event of March-April 2012}

\author{
D.~J.~Andrews,\altaffilmark{1}
S.~Barabash,\altaffilmark{2}
N.~J.~T.~Edberg,\altaffilmark{1}
D.~A.~Gurnett,\altaffilmark{3}
B.~E.~S.~Hall,\altaffilmark{4}
M.~Holmstr\"{o}m,\altaffilmark{2}
M.~Lester,\altaffilmark{4}
D.~D.~Morgan,\altaffilmark{3}
H.~J.~Opgenoorth,\altaffilmark{1}
R.~Ramstad,\altaffilmark{2}
B.~Sanchez-Cano,\altaffilmark{4}
M.~Way,\altaffilmark{5,6}
and O.~Witasse.\altaffilmark{7}\\
\vspace{11pt}
Corresponding author: David J. Andrews, Swedish Institute of Space Physics (Uppsala), Box 537, Uppsala 75121, Sweden.\\
 david.andrews@irfu.se
}

\date{March 2016. Preprint accepted for publication in \\J. Geophys. Res. (Space Physics).}

\begin{document}
\onecolumn

\maketitle

\begin{abstract}
We present initial analysis and conclusions from plasma observations made during the reported ``Mars plume event'' of March - April 2012.
During this period, multiple independent amateur observers detected a localized, high-altitude ``plume'' over the Martian dawn terminator [Sanchez-Lavega et al., Nature, 2015, doi:10.1038/nature14162], the cause of which remains to be explained.
The estimated brightness of the plume exceeds that expected for auroral emissions, and its projected altitude greatly exceeds that at which clouds are expected to form.
We report on in-situ measurements of ionospheric plasma density and solar wind parameters throughout this interval made by Mars Express, obtained over the same surface region, but at the opposing terminator.
Measurements in the ionosphere at the corresponding location frequently show a disturbed structure, though this is not atypical for such regions with intense crustal magnetic fields.
We tentatively conclude that the formation and/or transport of this plume to the altitudes where it was observed could be due in part to the result of a large interplanetary coronal mass ejection (ICME) encountering the Martian system.
Interestingly, we note that the only similar plume detection in May 1997 may also have been associated with a large ICME impact at Mars.

Key points:
\begin{itemize}
\item Data from ASPERA-3 and MARSIS were obtained during the unusual atmospheric plume event.
\item Plume observations likely follow the impact of large interplanetary coronal mass ejections.
\item Ionospheric plasma structures associated with the plume are not conclusively observed.
\end{itemize}
\end{abstract}
\twocolumn

\section{Introduction}

Mars, including specifically its surface, atmosphere, and induced magnetosphere, has been the subject of continuous in-situ study for nearly two decades.
Recent reports of remote observations of an extremely high-altitude `plume' were therefore something of a surprise~\citep[][hereafter \emph{SL15}]{sanchez-lavega15a}.
Over the interval of 12 March - 17 April 2012 following apparent opposition, observations made of Mars in the optical band by amateur astronomers from several distinct geographical locations showed the presence of a detached, bright feature above the Martian surface.
The feature was centered near 43$^\circ$ South latitude, 197$^\circ$ West longitude.
While projection effects make the determination of a `true' altitude almost impossible, conservative estimates indicate that the feature was present at altitudes up to $\sim$280~km, and extended over $\sim$11$^\circ$ of latitude.
Its longitudinal extent was inferred to be $\sim$11$^\circ$, and in several cases it was clearly observed to be rotating with the planet.
Importantly, in each instance the plume was found at the same location over the planet's surface within the uncertainties of the observations, and was only visible as that region traversed the dawn terminator.
Possible observations of the plume while it lay over the sun-lit disc of the planet were likely prevented by the bright surface below, and it was not observed as it crossed the dusk limb (the dusk terminator not being visible from Earth at this time).
Therefore, any diurnal variation of the plume is not constrained by the available observations~[\emph{SL15}].

\emph{SL15} explored several possible physical explanations for the observed plume, specifically the local condensation of water or CO$_2$ ice, atmospherically suspended dust, and auroral emissions.
Detailed observations of clouds at Mars have been extensively reported in the literature~\citep[see e.g.][]{montmessin07a, maattanen10a, gonzalez-galindo11a, maattanen13a}
The events of March-April 2012 appear to be of a very different class as the observed altitude of the plume is significantly higher than those at which CO$_2$ and/or H$_2$O are expected to be able to condense within the Martian atmosphere~[\emph{SL15}].
Measurements by Mars Climate Sounder on Mars Reconnaissance Orbiter have shown that water ice clouds may be present at higher altitudes than previously expected, i.e.\ up to $\sim$50~km~\citep{heavens10b} and have been shown to vary seasonally, diurnally, and in response to orographic forcing.
However, these observations are still made well below the altitudes we concern ourselves with in this paper.

Meanwhile, dust plays an important role in the dynamics of the Martian atmosphere, both at low and middle altitudes~\citep[e.g.][]{mccleese10a}.
The effects of global dust storms have been shown to be measurable up to ionospheric altitudes~\citep[e.g.][]{lillis08d, england12a, liemohn12a}.
Furthermore, the increased thermospheric mass densities and correspondingly increased photoelectron fluxes at a given altitude may persist even after the lower-altitude dust storm has subsided~\citep{xu14a}.
Distinct layers of dust may be present in the Martian atmosphere up to altitudes of $\sim$70~km, perhaps as the result of vertical transport due to thermal updrafts generated in regions of topographical variations~\citep[e.g.][]{guzewich13a, heavens15a}.
However, lofted dusty material has not hitherto been reported at the altitudes corresponding to the feature observed by \emph{SL15}, and~\cite{kleinbohl15a} have recently shown a lack of a long-lived dust layer in the middle atmosphere.
A clear body of evidence therefore exists for complex coupling between the behavior of dust and aerosols throughout the Martian atmosphere from the surface to the thermosphere, though the precise details of much of this coupling remain to be understood.

\emph{SL15} also briefly explored the possibility that the observed features were the manifestation in the optical band of a localized auroral emission.
The surface location of these observations is consistent with that reported previously for Martian aurora~\citep{bertaux05a, gerard15a}, being over a region of intense and highly structured crustal magnetic fields~\citep{acuna99a}.
However, if the plume was in fact an auroral emission, its brightness would vastly exceed the spectral observations made by the UV spectrometer on Mars Express (MEX), by at least 3 orders of magnitude, making it significantly brighter than any auroral emission observed at Earth or indeed any other planet.

A further possibility, not discussed by \emph{SL15}, is the formation of this layer by ablation of material from a meteor~\citep[e.g.][]{molina-cuberos03a}.
Ionized material from such impacts has a long lifetime at ionospheric altitudes, and has recently been observed by NASA's MAVEN mission following the close approach of comet Siding-Spring~\citep{schneider15a}.
The influx of material from Siding Spring was clearly associated with the formation of a layer of ionized magnesium in the Martian atmosphere, with peak densities at altitudes of $\sim120$~km.

In summary, our current understanding of the Martian atmosphere does not include processes that can act to form the observed high altitude plume reported by $\emph{SL15}$.
In this paper we concentrate on coincident in-situ and remote plasma observations made during this period by MEX.
The main layer of the Martian ionosphere is formed through photoionization of CO$_2$, and has its peak in density at altitudes of $\sim$135~km at the sub-solar point, rising to $\sim$180~km at the terminator~\citep[e.g.][]{morgan08a}.
The high projected altitude of the plume would therefore place it well above the main peak of the ionosphere, in the region where draped heliospheric fields typically dominate the magnetic field configuration.
During the interval the plume was observed, the second of three measurement campaigns organized in part by the Mars Upper Atmosphere Network (MUAN) was underway, spanning the March - April period~(see details given by~\cite{opgenoorth13a}).
The coincidence in time is not surprising, since the MUAN campaign was conducted following apparent opposition, when solar wind measurements made by dedicated spacecraft at Earth could be most reliably extrapolated to the orbit of Mars, yielding the best-possible measurements of the upstream solar wind at Mars.
During this interval, MEX made several passages over the surface region of Mars where this plume was observed in observations made from Earth.
However, the phasing of the MEX orbit was such that this surface region was crossed at the opposite terminator: the plume was observed from Earth over the dawn terminator, while the MEX data studied here were obtained at dusk.
This local time offset prevents our ability to study the plume directly, as we have no information about the persistence of the plume over a full rotation of Mars.
The observations presented here nevertheless provide relevant information about the state of the ionosphere in this region, and any diurnal variation present.
Additionally, we also report on the state of the solar wind during this interval, and find at least a tentative correlation between the plume observations and the preceding impact of solar wind shocks at Mars, associated with interplanetary coronal mass ejections (ICMEs).

The response of the Mars ionosphere and induced magnetosphere to ICME events has long been studied~\citep[e.g.][]{crider05a, edberg10a, opgenoorth13a, morgan14a, jakosky15b}.
The general consensus is the enhanced dynamic pressure associated with these events leads to short-term increases in the rate of atmospheric escape, along with a compression of the plasma boundaries that separate the ionosphere from the upstream solar wind.
The precise mechanisms by which momentum is transferred from the solar wind to the escaping planetary ions remain the subject of detailed study, as does their relative importance.
Similarly, solar flares and the associated increase in ionizing UV light has been shown to enhance the Martian ionosphere~\citep[e.g.][]{mendillo06a, mahajan90a}, as has the precipitation of shock-accelerated solar energetic particles (SEPs) into the atmosphere~\citep[e.g.][]{lillis12a, ulusen12a, nemec14a}.
It must be noted, however, that these three sources of short-term variations in the ionosphere and induced magnetosphere, while themselves often having a common root cause on the solar surface, often are incident at Mars at markedly different times, as the flare naturally travels at the speed of light, solar energetic particles some significant fraction thereof, but following heliospheric magnetic field lines, and the bulk of the ICME $\sim$1-4 days later, depending on the shape and propagation speed of the ejecta.
The presence of a flare, SEP flux increase, or an ICME is not \emph{a priori} a reliable predictor of the others, either concurrently or shortly afterwards.
Distributed multi-point measurements and/or advanced modeling schemes are required to fully understand the causal relationships between these observations~\citep[e.g.][]{falkenberg11a}.

\section{Instrumentation and Models}
The periapsis altitude of MEX during the period in which the plume was observed was $\sim$335~km, i.e.\ somewhat above the uppermost altitudes at which the phenomenon was observed.
Extended series of measurements were taken both with the Analyzer for Space Plasmas and Energetic Atoms (ASPERA-3) plasma instrument suite~\citep{barabash06a} and the Mars Advanced Radar for Sub-Surface and Ionospheric Sounding (MARSIS)~\citep{picardi04b, gurnett05a}, which we discuss in detail in this paper.

ASPERA-3 comprises a suite of sensors dedicated to the measurement of ions, electrons and energetic neutral atoms.
In this paper we use data from the Ion Mass Analyzer (IMA) sensor, which determines the energy and mass per charge of incident ions, and has a field of view of $\pm$45$\times$360$^\circ$ (elevation $\times$ azimuth), using electrostatic deflectors to cover the elevation angle.
The instrument is capable of separately resolving H$^+$, He$^{++}$, O$^+$, O$_2^+$ and CO$_2^+$ in the energy range 0.01-36~keV per charge.
A full scan of mass, energy, azimuth and elevation is completed every 192~s.
From this, the bulk moments of the plasma may be numerically computed yielding density, velocity and temperature (under ideal conditions).
In practice, part of the instrument's field of view can be obscured by the spacecraft bus and solar arrays, and spacecraft potential variations can limit its ability to measure cold ionospheric plasma flows.

We note that ASPERA-3 is not a dedicated solar wind monitor and owing to the orbit of MEX cannot continuously sample the solar wind.
The fraction of each $\sim$7~h orbit for which MEX is in the undisturbed solar wind varies significantly, but is typically not more than $\sim$75\%, and often significantly less than this (occasionally falling to near zero, when apoapsis is located in the Martian induced magnetotail).
During the period specifically studied here, MEX spends $\sim$4~h per orbit in the solar wind.
We therefore supplement these discontinuous in-situ solar wind measurements with higher-precision and continuous measurements made by dedicated spacecraft at Earth orbit, specifically the Advanced Composition Explorer (ACE).
These measurements are extrapolated to Mars orbit using the 1-D MHD Michigan Solar Wind Model (MSWIM).
Full details of this approach, along with an evaluation of its effectiveness are given by~\cite{zieger08a}.
Briefly, plasma moments measured by ACE are transformed into an inertial frame, and used as time-dependent boundary conditions in simulating the solar wind stream as it evolves to Mars's orbital location.
During the period of the Martian plume observations, this propagation is expected to be most reliable, as both the radial distance and the angular separation between Earth and Mars have their smallest values.

MARSIS comprises a 40~m tip-to-tip dipole and associated electronics required to send and receive radio pulses.
The instrument is operated at periapsis, at altitudes typically below $\sim$1200~km.
One of the key aspects of the MUAN campaign run during the period studied here is the generally enhanced volume of data taken with MARSIS in Active Ionospheric Sounding (AIS) mode.
When operated in AIS mode, the instrument transmits a short pulse of $\sim$100~$\mu s$ duration at a given frequency $f$, before ``listening'' for reflections of the pulse from the Martian ionosphere below at the same frequency.
The time delay between the transmission of the pulse and any detected reflection gives the distance to the reflection site.
The process is repeated at 160 logarithmically spaced pulse frequencies from $\sim$0.1 to 5.5 MHz, forming a so-called ``ionogram'', with time delay and frequency as its two axes.
The variation in the curve of the time delay to the ionospheric reflection with frequency can be numerically inverted to yield a profile of ionospheric plasma density with altitude, from the spacecraft down to the ionospheric peak density (below which, all pulses instead propagate through the ionosphere and reflect from the surface of the planet).
Full details of this inversion process as applied to MARSIS data are given by~\cite{morgan13a}.
Finally, we also note that the relatively high-power of the transmitted pulse is sufficient to disturb the plasma around the antenna in a complex fashion, giving rise to distortions in the data at the local plasma frequency, as well as ``pseudo-echoes'' at the local electron gyroperiod in regions where the magnetic field is strong, typically above $\sim$20~nT~\citep{gurnett05a}.

In the following section, we briefly report on the content of these plasma observations, the state of the solar wind, and the tentative conclusions we draw from comparisons with the timings of the reported optical observations.

\section{2012 March - April Observations with MEX}

\begin{figure*}[tp]
\includegraphics[width=0.95\linewidth]{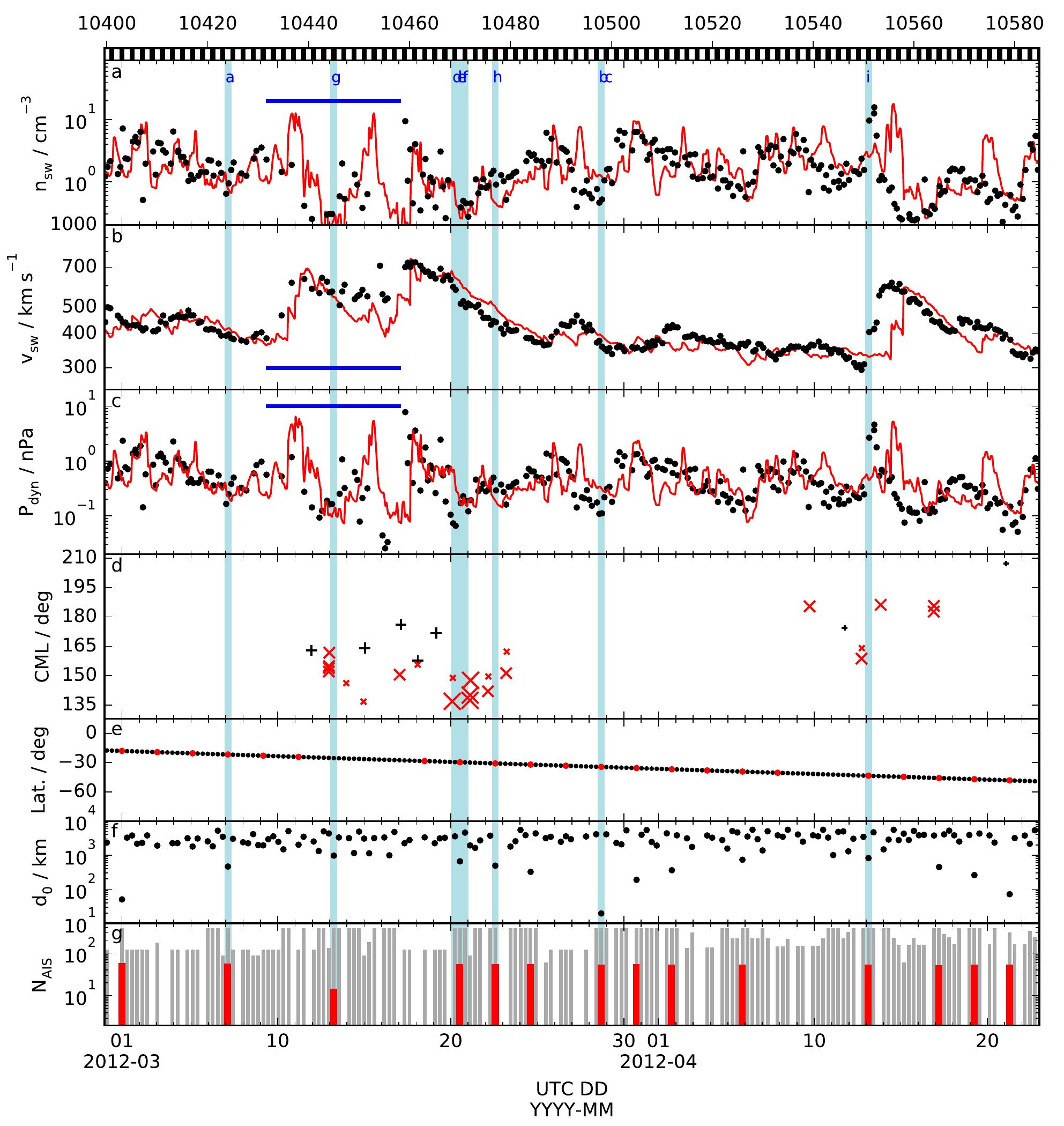}

\caption{Summary plot of plasma data obtained during the two plume events.
	a) Solar wind proton density $n_{sw}$ as measured by ASPERA-3/IMA when outside the Martian bow shock (black circles), and simulated values from MSWIM(red line).  The horizontal blue line indicates the interval in which IMA measurements are degraded by disturbed solar wind.
	b, c) As (a), but showing solar wind bulk velocity $v_{sw}$ and dynamic pressure P$_{dyn}$.
	d) Timeline of the optical observations of the plume, showing the central meridian longitude (CML) of each observation. Positive detections of the plume are plotted as red `$\times$' symbols, with size from small to large indicating the assessed quality of the observation.  None-detections of the plume are shown as similarly coded black `$+$' symbols.
	e) MEX latitude at periapsis, highlighted red when periapsis occurs at longitudes close to the plume location.
	f) Approximate closest surface distance of each orbit to the plume nominal center, only recorded when AIS is operating.
	g) Number of MARSIS/AIS soundings performed per orbit of MEX (grey bars). Red smaller bars indicate the number of soundings performed over the plume location.
	The alternate black and white blocks on the upper edge of the figure show the orbits of MEX (numbered every 20).
	Light blue vertical bars indicate those orbits from which we report MEX data in subsequent figures.}
	\label{fig:obs}
\end{figure*}

Figure~\ref{fig:obs} summarizes the various plasma observations made during the interval identified by~\emph{SL15}, along with the periapsis locations of MEX, and the timing of individual plume observations.
Throughout this period, the azimuthal separation between Mars and Earth varied from $\sim5^\circ$ (Mars leading) to $\sim30^\circ$ (Earth leading), with radial alignment occurring on 5 March 2012.
Throughout all panels in Figure~\ref{fig:obs}, we highlight in light blue those MEX orbits for which we will later show individual measurements made with MARSIS.
Panels a to c of Figure~\ref{fig:obs} show respectively measured solar wind density $n_{sw}$, speed $v_{sw}$ and dynamic pressure P$_{dyn}$, obtained from ASPERA-3 ion measurements while MEX was in the solar wind (black circles).
From $\sim$12:00 (UTC) on 9 to $\sim$00:00 on 18 March the quality of these ASPERA-3 data are significantly reduced, almost certainly due to the impact of at least one very large interplanetary coronal mass ejection (ICME) on the Martian induced magnetosphere.
The associated penetrating radiation and enhanced particle fluxes overloaded an internal buffer in the instrument, and large volumes of data were irretrievably lost.
Furthermore, extreme rarefactions in solar wind density in the wake of the ICME pose further instrumental problems for the measurements of these bulk parameters.
This interval in which the instrument performance is degraded is marked by the horizontal blue bars.
However, we are able to supplement these measurements through the use of the results of the MSWIM propagation, as shown by the red traces in panels a-c of Figure~\ref{fig:obs}.
These ACE-derived estimates of the solar wind at Mars corroborate the impact on Mars of a major ICME with peak velocities exceeding $\sim700$~km~s$^{-1}$ on $\sim$9 March.
In addition, one or possibly two subsequent smaller ICMEs, launched on 9 March, are predicted to impact Mars beteen 12 and 14 March.
Large density depletions are found in the wakes of this chain of ICMEs, with densities falling below values that can be meaningfully interpreted either in ASPERA measurements or MHD simulation results.
Overall, the ICME that was launched from the Sun on 7 March 2012 and impacted Earth on 9 March was a significant event, and likely one of the most intense to hit Mars during the ascending phase of solar cycle 24.

Outside of this disturbed interval, the agreement between the ASPERA-3 measured solar wind parameters and those propagated from Earth using MSWIM is in general reasonable, particularly in terms of expected velocities.
A further significant shock appears to arrive on April 13 with a large rise in solar wind density, followed by a doubling of the solar wind velocity.
This perturbation is more characteristic of a corotating stream interaction region (SIR), and is well accounted for in the MSWIM estimates, albeit with a small delay of $\sim$1 day.
Such SIR fronts, which have a distinct ``sawtooth'' density profile are found at the interfaces between slow and fast solar wind streams, and can often persist throughout several solar rotations.

Panel d of Figure~\ref{fig:obs} shows the timings of the ground-based telescope observations made from Earth in which the Mars plume was detected.
The longitude of the sub-Earth point on Mars at the time of each observation is plotted, commonly referred to as the ``central meridian longitude'' (CML) [Sanchez-Lavega, private communication].
Precise calculation of this quantity requires reliable information regarding the timing of each observation, which is available for almost all the observations noted by $\emph{SL15}$.
Small, medium and larger red `$\times$' symbols signify ``tentative'', ``clear'', and ``excellent'' quality positive detections of the plume.
Equivalently sized black `$+$' symbols represent corresponding quality non-detections, i.e. successful observations of Mars which did not show a plume, but had the required resolution to be able to resolve one were it present at the terminator.
We ascribe more weight to those observations in which the plume was first noted, on 20 and 21 March, in contrast to those which were retrospectively found in re-examined data.
We also add weight to the observations where the image quality was particularly clear, or multiple detections were made on a single night, or an image sequence was obtained showing the motion of the plume over the limb.
Conclusive statements regarding the presence or absence of the plume can be made only intermittently with the available observations.
We note that the clearest and most frequent plume sightings (large red `$\times$' symbols indicating positive detections) all occur within the first event, around 20 March, while the event may begin as early as March 13 following the first clear detection.
A localized feature rotating over Mars' surface will only be visible in a narrow range of CML.
For the initial observations in March, positive detections only occur with CML less than $\sim160^\circ$, while the non detections are all made at larger values.
Hence, the interspersed non-detections of the plume are likely not indicative of its absence, and instead may only be the result of unfavorable viewing geometry.
Little can be safely concluded regarding the plume activity, or lack thereof in the interval March 23 to April 10, before it is once again observed for a period of $\sim$7 days until April 17.
In our assessment, only clear non-detections in the same CML range where the plume was initially seen can yield firm constraints on its duration.
Since these are lacking from the available observations, the extent in time for which the plume was present in the Martian atmosphere cannot be properly constrained.

In the final three panels of Figure~\ref{fig:obs} we show parameters regarding the orbit and operation of MEX during this interval.
Again, we note that the local time (LT) of periapsis of MEX during these observations varied steadily between 17:20 - 17:50~h LT throughout this period, and therefore provide a dusk counterpart to the optical observations at the dawn terminator.
Panel e of Figure~\ref{fig:obs} shows the latitude of periapsis of MEX, slowly decreasing through the southern hemisphere with time.
Red markers highlight those periapses at longitudes close to the plume nominal center, specifically 175$^\circ$ - 220$^\circ$ West (or 140$^\circ$ - 185$^\circ$ East, for commonality with other MEX publications).
Panel f of Figure~\ref{fig:obs} shows the minimum surface distance between the spacecraft and the plume nominal center on each orbit.
The closest approach to the average plume surface location is within 20~km, occurring on orbit 10498 on 28 March.
This likely places MEX (at dusk) directly above the region where the plume was seen to be active (at dawn) at this time, given the extended horizontal size of the plume.
Finally, in panel g of Figure~\ref{fig:obs} the grey bars show the number of MARSIS AIS soundings performed at periapsis of each orbit, with those shown in red indicating the number of soundings made over the plume region.
Specifically we define ``over the plume region'' as being latitude -43.1$\pm10.8^\circ$ and longitude 197.1$\pm22.2^\circ$ (West), where we note that we have taken the `extreme' range given by \emph{SL15}, and further doubled the longitudinal extent.
Line of sight projection effects can introduce significant ambiguities in both spatial location and altitude, and modestly increasing the longitudinal extent of the region of interest, making it of approximately equal spatial extent in both the zonal and meridional directions seems reasonable to us.

Summarizing the data shown in Figure~\ref{fig:obs}, we see that the clearest plume detections on 20 and 21 March follow the impact of a major ICME, along with one or two smaller trailing ICMEs and the disturbed solar wind in their wakes.
The impact of at least the first large ICME, and associated energetic particles is confirmed in in-situ measurements from ASPERA-3.
A smaller solar wind enhancement is also present in the second cluster of plume observations around April 14, though this likely to be a SIR rather than an ICME.
The apparent positive plume detection occurring on 9 April 2012 does not show any immediately preceding solar wind enhancement according to in-situ measurements made with ASPERA-3 (Figure~\ref{fig:obs}, panels a-c).
The lack of recorded non-detections in the period March 24 to April 9 is unfortunate, as it prevents us making firm statements about the duration of the major plume event following the large March 9 ICME.
We conclude that it is at least possible that the second series of plume observations is simply a direct continuation of the first.

The lack of a magnetometer onboard MEX prevents measurements of solar wind convection electric field direction, which exerts significant influence over the configuration of the Martian ionosphere and induced magnetosphere~\citep[e.g.][]{dubinin06a, brain06c}.
While estimating the orientation of the upstream magnetic field from the MSWIM propagations is possible, significant deviations can be expected due to evolution of the solar wind, particularly in response to the ICMEs embedded within it.
We therefore do not show these data, but instead only briefly comment that there is very weak evidence to suggest that more of the individual plume observations are associated with a ``toward'' configuration of the Parker spiral than the opposite ``away'' configuration.

\begin{figure}[tp]
	\includegraphics[width=0.9\linewidth]{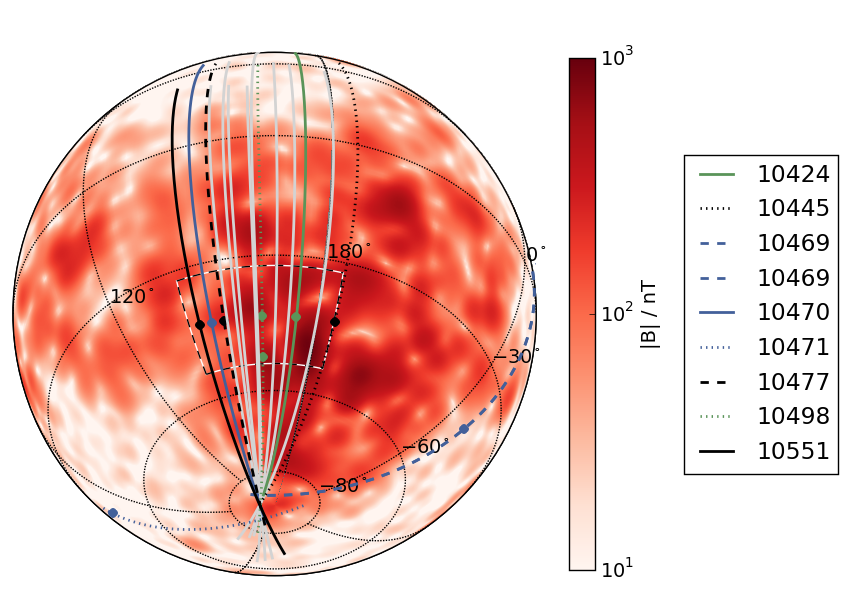}
	\caption{Projected ground tracks of MEX over the plume location (black and white dashed box). Grey lines indicate all those orbits entering the region of interest during the interval shown in Figure 1.  Colored and labelled lines indicate those orbits discussed in the paper.  The surface of the planet is colored according to the strength of the crustal magnetic field according to the model of~\cite{lillis10b} at 150~km altitude.}
	\label{fig:pos}
\end{figure}

Figure~\ref{fig:pos} shows the trajectory of MEX projected onto the surface of Mars, where the surface is shown color-coded according to the crustal magnetic field intensity $|B_{Crustal}|$ using the model of~\cite{lillis10b} evaluated at an altitude of 150~km.
Colored trajectories indicate those orbits highlighted in Figure~\ref{fig:obs}, from which data are later shown, according to the label on the right.
Other orbits which pass through the plume region of interest bounded by the black and white dashed line during March and April 2012 are shown grey.
In each case, only the periapsis segments are shown, corresponding to the periods when MARSIS is operating in AIS mode.

\begin{figure*}[tp]
	\includegraphics[width=0.9\linewidth]{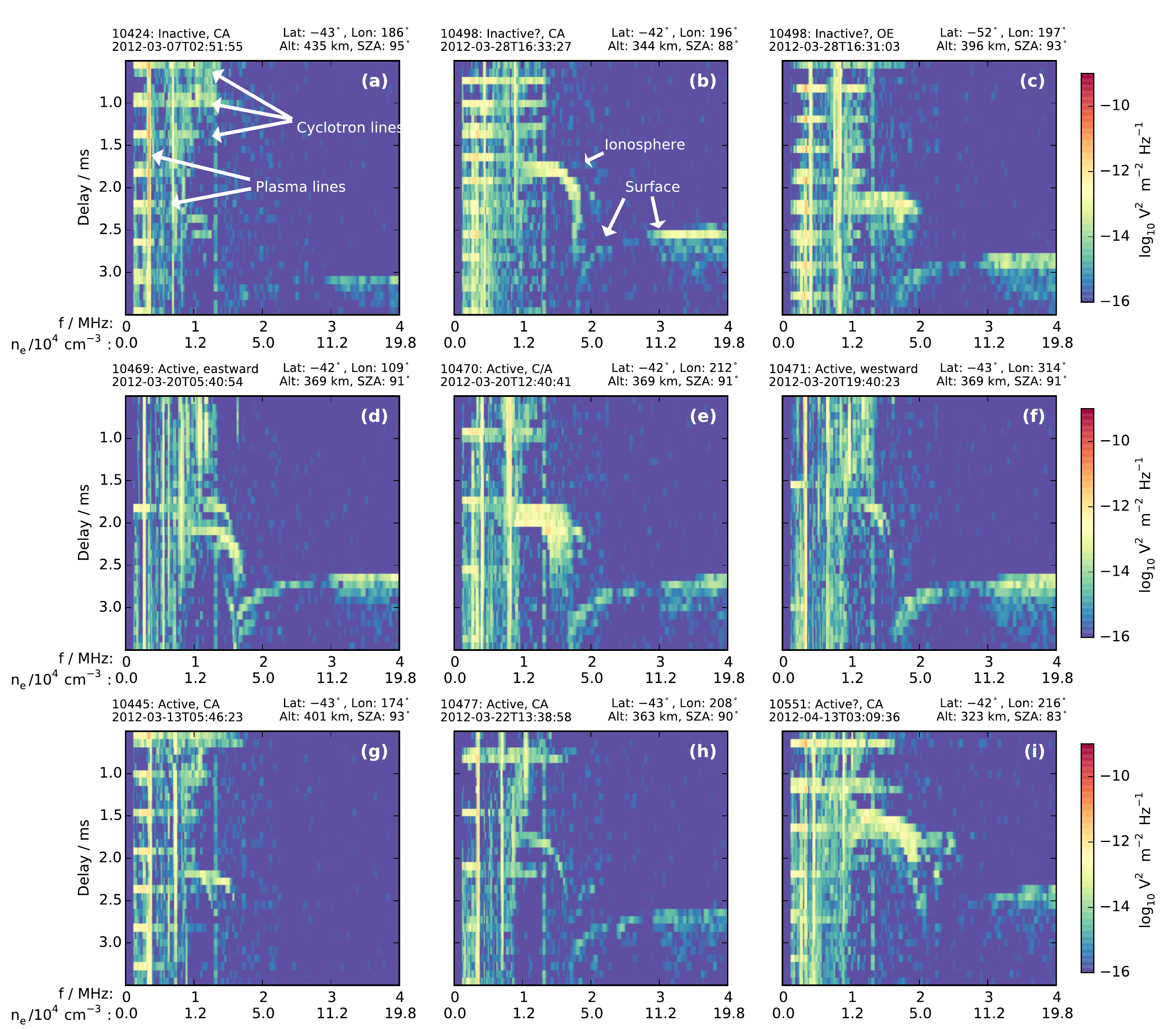}
	\caption{
		A selection of 9 ionograms obtained with MARSIS during the interval shown in Figure~\ref{fig:obs}.  Received signal on the antenna is color-coded versus delay time and sounding frequency (equivalent densities are also indicated on the x-axes).
		The orbit number, description, and UTC time are indicated in the upper left of each sub-panel, along with the position of MEX in the upper right.
		For presentation purposes, these plots are not shown in time-order.
	}
	\label{fig:ionograms}
\end{figure*}

In Figure~\ref{fig:ionograms} we show ionograms obtained with MARSIS/AIS at several instances both before, during, and after individual plume observations were made by \emph{SL15}.
Each of the 9 panels shows an individual ionogram, obtained at the orbit and time indicated in the upper left of the panel.
The projected locations of these ionograms are shown by the appropriately colored circles on the mapped trajectories in Figure~\ref{fig:pos}.
Each ionogram shows the color-coded signal intensity measured on the antenna versus delay time (y-axis) and transmitted frequency (x-axis, with equivalent plasma density also indicated).
Characteristic features are labelled in Figure~\ref{fig:ionograms}a and b.
These are namely the vertical plasma lines occurring at integer multiples of the local electron plasma frequency $f_{pe} = \sqrt{n_e e^2 / \epsilon_0 m_e 4 \pi^2}$ surrounding the spacecraft, horizontal cyclotron lines occurring at multiples at the electron gyroperiod $\tau_{ce} = qB/2\pi m_e$, the ionospheric reflection trace extending to larger delays at higher frequencies as the peak density is approached.
Finally the surface reflection of radio waves is visible at the highest frequencies, which pass completely through the ionosphere.
Interpretation of these ionograms is not without its subtleties, and we refer the reader in particular to related papers by~\cite{gurnett05a, duru06a, morgan08a} and \cite{morgan13a} for further details.

Firstly, we report those MARSIS/AIS observations shown in Figure~\ref{fig:ionograms}a-c, made during intervals when the plume was not visible according to the amateur observations from Earth.
Figure~\ref{fig:ionograms}a shows the ionogram obtained on 2012 March~07 over the dusk terminator, closest to the site where the plume would later be detected from March 13 onwards at the dawn terminator.
The characteristics of this ionogram are essentially unremarkable - a very restricted reflection at $\sim$1~MHz is observed at $\sim$2.4~ms delays, likely ionospheric in origin.
The solar zenith angle (SZA, 0$^\circ$ at the sub-solar point, 90$^\circ$ at the terminator at the surface) of this observation puts the spacecraft behind the geometric terminator.
However, the vertically extended ionosphere remains sun-lit to an SZA of $\sim$110$^\circ$, and hence the presence of ionospheric plasma sub-spacecraft in this location is not unusual.
Figure~\ref{fig:ionograms}b and~\ref{fig:ionograms}c meanwhile show observations obtained on March 28, following the first series of confirmed plume detections ending on March 23.
Figure~\ref{fig:ionograms}b shows the ionogram obtained closest to the planetographic location of the plume nominal center, which again shows a fairly unremarkable ionospheric trace indicating a stratified ionosphere, along with cyclotron lines that are more closely spaced, indicating a more intense magnetic field at the spacecraft than shown in Figure~\ref{fig:ionograms}a.
We also note that the observation shown in Figure~\ref{fig:ionograms}b is the closest obtained to the plume nominal center throughout the period studied.
Figure~\ref{fig:ionograms}c was obtained $\sim$10$^\circ$ further south of Figure~\ref{fig:ionograms}b on the same orbit, closer to the terminator, and shows a reflection at higher peak frequency (and therefore density).
The ionospheric trace is also ``thicker'', extending over a larger range of delay bins within the instrument, possibly indicating a more disturbed ionosphere, with more horizontal irregularities giving rise to multiple reflection sites.
Furthermore, the ionospheric trace now overlaps with that of the ground, an effect which is only possible when the ionospheric reflection is at least in part being received from an off-nadir direction, i.e. at oblique incidence~\citep{duru10b}.
This provides further evidence for a large degree of structuring of the ionosphere at this location.

We note that it is in principal possible to invert these reflections, accounting for the dispersion of the radio waves during their passage to and from the reflection point at each frequency and yielding a profile of electron density versus altitude above the surface~\citep[see e.g.][]{morgan13a}.
However, we do not perform this operation on these data, for several reasons.
Firstly, owing to the location of these soundings near the terminator, the horizontal structuring of the ionosphere makes it highly likely that distortions will be present in the final results of such an inversion.
Principally, this will lead to underestimations of the true altitude of particular features, but also can cause further distortions, smearing out any real extra layers that may be present.
Secondly, the lower density of the ionosphere at this location as compared e.g. to the sub-solar ionosphere, means that the interpolation from the spacecraft to the lowest frequency ionospheric reflection represents a large fraction of the total trace.

Moving now to Figure~\ref{fig:ionograms}d-f, we show a sequence of 3 successive orbits 10469-10471 (all taking place on March 20).
The central ionogram Figure~\ref{fig:ionograms}e shows the observation made closest to the plume center, a matter of hours before it would then be observed as the same region of Mars surface traverse the dusk limb.
Figure~\ref{fig:ionograms}d and~\ref{fig:ionograms}f then show the observation made at the same latitude and SZA on the preceding and following orbits, but at spacecraft longitudes rotated $\sim$100$^\circ$ eastward and westward of the plume center, respectively.
The trace observed closest to the plume in Figure~\ref{fig:ionograms}e displays a much thicker reflection, than those before (\ref{fig:ionograms}d) and after (\ref{fig:ionograms}f), again indicating small-scale structuring of the ionosphere in this location.
The ionospheric reflection in Figure~\ref{fig:ionograms}d comprises multiple individual traces, indicating oblique reflections from more distant points in the ionosphere, away from the nadir direction, while that in Figure~\ref{fig:ionograms}f is fainter, but otherwise rather unremarkable.
The identical illumination conditions of the ionosphere during these three ionograms is at odds with their varied presentation.
However, we cannot confidently ascribe any of this variation to the presence of the plume in Figure~\ref{fig:ionograms}e, or the lack thereof in Figure~\ref{fig:ionograms}d and~\ref{fig:ionograms}f.
All of the variations seen in these three ionograms could easily be ascribed instead to the different crustal field conditions present between these locations.
Stronger crustal fields are clearly detected in Figure~\ref{fig:ionograms}e, owing to the closer-spacing of the cyclotron lines relative to Figure~\ref{fig:ionograms}d and~\ref{fig:ionograms}f.
Specifically, the modeled crustal field strength at 150~km altitude at the location of the ionogram shown in Figure~\ref{fig:ionograms}e is $\sim$100 nT, its orientation is radially outward from the planet's surface~\citep{lillis10b}.
In contrast, modeled crustal fields at the locations of Figure~\ref{fig:ionograms}d and~\ref{fig:ionograms}f are much weaker $\sim$10~nT or less, and therefore negligible compared with typical draped magnetic field intensities.
The ionosphere in regions of intense near-radial crustal fields is well known to be elevated with respect to other regions~\citep[e.g.][]{gurnett05a}, and often displays such a ``thick'' reflection trace.

Finally, Figure~\ref{fig:ionograms}g-i show observations made on March 13, 22 and April 13, respectively, all made when the plume was reported to be active by \emph{SL15}, on the closest approach to the plume location on each orbit.
Clear, and varied, ionospheric reflections are present in each case.
Multiple reflections are present in both Figure~\ref{fig:ionograms}g and i, while h shows a single trace.
A broader range is seen in the peak frequency of the ionosphere, and therefore its density in these three examples than the 6 discussed previously.
A third example of a thicker reflection can be seen in Figure~\ref{fig:ionograms}i, the only plot we show from the second run of plume observations, made during April 2012.
No surface reflection is evident in Figure~\ref{fig:ionograms}g, likely indicating the presence of a plasma layer at altitudes below the nominal ionospheric peak density, in which collisional absorption of the sounding pulse occurs before the surface is reached.
Such effects have been studied previously by~\cite{morgan10a} and~\cite{witasse01a}, and have been related to the precipitation of high energy particles into the atmosphere, causing low-altitude ionization.

\begin{figure*}[t]
	\includegraphics[width=0.9\linewidth]{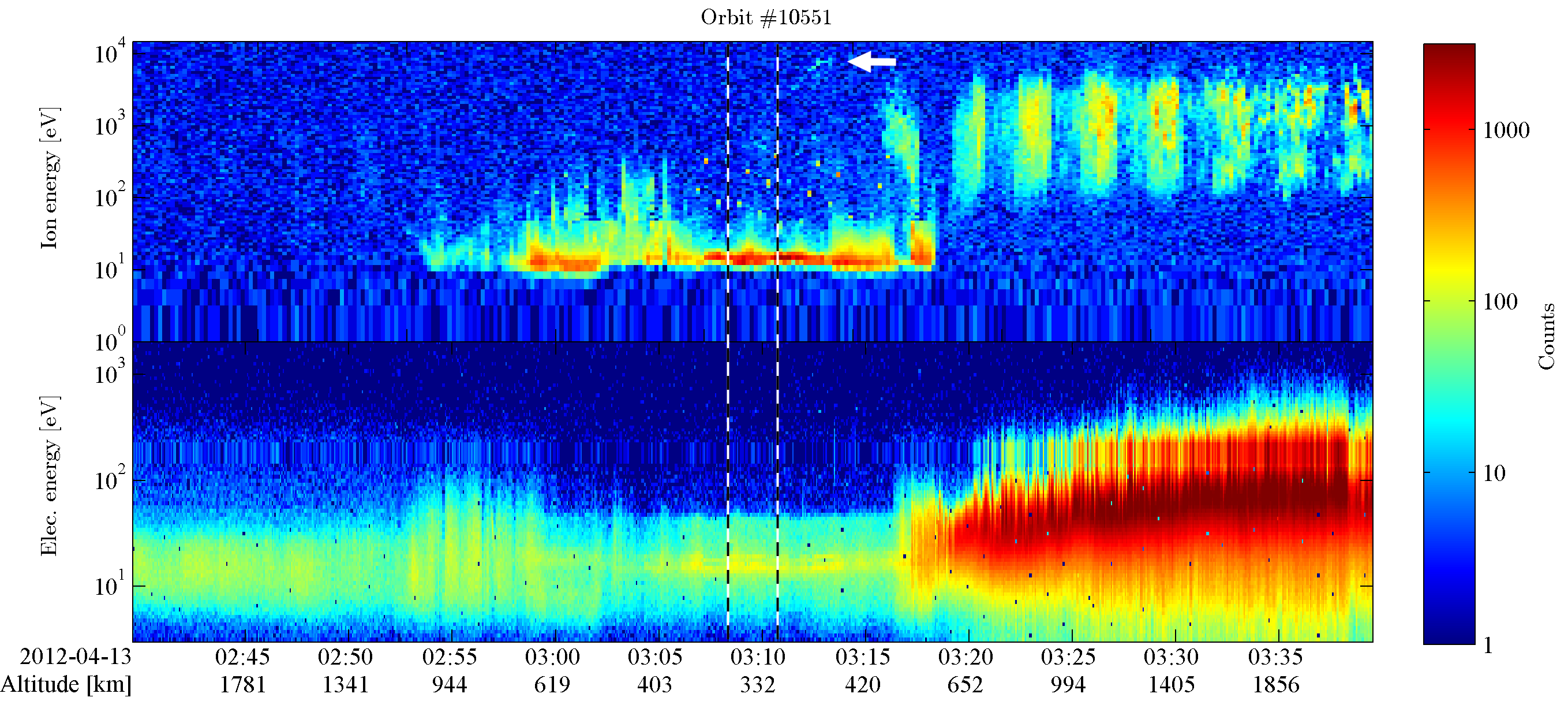}
	\caption{ASPERA-3 IMA (upper panel) and ELS (lower panel) data obtained on MEX orbit 10551, 13~April~2012, coincident with the second series of individual plume observations. Count rates for ions and electrons are shown color-coded, summed over all anodes and scan directions for each sensor, respectively.
	Vertical dashed lines bound the interval for which MEX was above the plume location, i.e. within the highlighted box shown in Figure~\ref{fig:pos}.
	The white arrow indicates the accelerated ion feature discussed in the text.
	}
	\label{fig:asp}
\end{figure*}

In Figure~\ref{fig:asp} we plot spectra obtained by the ion and electron spectrometers of ASPERA-3 during orbit 10551 on 13 April 2012.
The duration of the passage of MEX through the plume location depicted in Figure~\ref{fig:pos} is marked by the vertical dashed black lines, and the closest approach to the plume center occurred at 03:10, coincident with the ionogram shown in Figure~\ref{fig:ionograms}i.
Very shortly after this, accelerated planetary ions were observed at unusually high energies, up to $\sim$7~keV, indicated in the spectra by the white arrow.
While the mass-resolving capabilities of IMA are not sufficient at these high energies to resolve the species, these are most likely O$^+$ or $\mathrm{O_2^+}$.
No associated signature is present in the electron spectra obtained at the same time.
Their high energy indicates a substantial acceleration of these ions has taken place, presumably from much lower energies characteristic of thermal ions in the Martian ionosphere.
Indeed, ionospheric heavy ions are simultaneously observed in the same time period as the energetic ions, indicating a mixed population.
Consideration of the look-direction of the IMA sensor during these observations suggests that these accelerated ions are traveling anti-sunward.
Taking the distance to the sub-solar bow shock point as an upper limit for the length scale over which the acceleration process could have acted, this would suggest a minimum accelerating (uniform, steady) electric field of $\sim$1~mV/m, directed anti-sunward.
Such an electric field would be required to accelerate a singly-charged planetary ion from rest to the observed energy.

\section{Discussion}
\subsection{Summary}

We have presented observations of the Martian ionosphere and induced magnetosphere obtained during the period March - April 2012, during which an anomalously high-altitude atmospheric `plume' was reported by \emph{SL15}.
In situ solar wind measurements were regularly obtained by MEX during this period, and several large ICMEs were observed to impact the Martian system, the largest of which arrived on $\sim$9 March, with a modest density enhancement and speeds exceeding $\sim$700 km$\;$s$^{-1}$.
Further confirmation of the arrival of the associated shocks, and following rarefaction regions was obtained using the MSWIM data-driven MHD simulation.
The first confirmed observations of the plume were then made on 13 March, with the clearest examples occurring later on 20 and 21 March, following this ICME impact and the wake-like structures in the solar wind that followed.
The 3-4 days that elapsed between the closest solar wind shock arrival and these clear observations do not support necessarily a direct connection between these phenomena.
However, the reported nondetections of the plume are essentially all consistent with observational restrictions arising from the CML at the time of observation, with the plume not being reliably observed during this March event for CMLs greater than $\sim$160$^\circ$.
Higher cadence and more continual observations would therefore have been required to reliably constrain the duration for which the plume was active, by ensuring that a broader range of CML was surveyed.
Making firm statements about both the start and end time of this plume event is not possible with the available optical observations.
While the majority of the plume observations made in April follow the impact of a reasonably strong SIR, the distinct possibility remains that the second set of observations may be a continuation of the first event.

We remind the reader that the MEX ionospheric observations presented here were obtained at local times close to the dusk terminator in each case, and therefore are almost exactly opposite to the dawn terminator region in which the plume was visible.
The lack of a clear signature in MARSIS/AIS soundings associated with the plume could therefore be construed as being consistent with at least some level of diurnal variation in the plume, either in altitude, horizontal extent, or its formation and dissipation on diurnal timescales.
In this context, we note that ionospheric density structures regularly seen by MARSIS in regions of intense crustal fields have been postulated to undergo systematic diurnal variation, forming and growing throughout their passage through the sunlit ionosphere, before dissipating on the nightside due to rapid ion-electron recombination~\citep{duru06a, andrews14a, dieval15a}.
However, despite the common altitude range of these phenomena, the rarity of observations of high-altitude atmospheric plumes contrasted with the very regular ionospheric oblique echo detections does not immediately suggest a causal relationship between these phenomena.

In summary, the events reported by \emph{SL15} are clearly interesting, and remain without explanation.
Both the plume's location in a region of intense crustal magnetic fields, and it's potentially interesting timing following a period of relatively extreme solar wind disturbances, and the ionospheric altitudes at which it was detected, collectively suggest that a direct connection is perhaps possible.
However, the available data during this event, and the wide separation in local time between observations made at Mars by MEX and the reported plume locations clearly limit the strength of the conclusions we may draw.
The ionospheric plasma density observations made by MARSIS over the plume region are best described as `typical' for that region of the Martian ionosphere, i.e. containing both elevated densities compared to other longitudes, localized density enhancements producing oblique echoes, and possible irregularities giving a dispersed reflection.

As was noted in the introduction, these MEX data were obtained during the second of three observation campaigns organized by the MUAN group, in each case at and following the apparent opposition of Mars, as this period provides the most reliable opportunity to extrapolate solar wind measurements made at Earth to Mars orbit.
No similar plume detections were reported during the other two campaigns, which took place during the spring of 2010 and 2014.
However, during neither of these intervals was a similarly extreme solar wind encountered as during the March 2012, as will be discussed later.

\subsection{Hubble Observations in May 1997}

In addition to the amateur ground-based observations reported by \emph{SL15}, they also conducted a search of Hubble Space Telescope (HST) observations of Mars, and noted a qualitatively similar plume-like feature in an observation made on 17 May 1997, from 17:27-17:41 UTC.
No relevant in-situ plasma measurements were available at Mars during this event (MGS would arrive later the same year, with only a very limited ability to resolve such disturbances in the solar wind).
However, our tentative conclusions that the formation of these plumes may in some way be related to the passage of strong solar wind disturbances is somewhat strengthened, as we note that a significant ICME was launched from the Sun on 12 May 1997 and impacted Earth on 15 May 1997.

\begin{figure}[t]
	\includegraphics[width=0.9\linewidth]{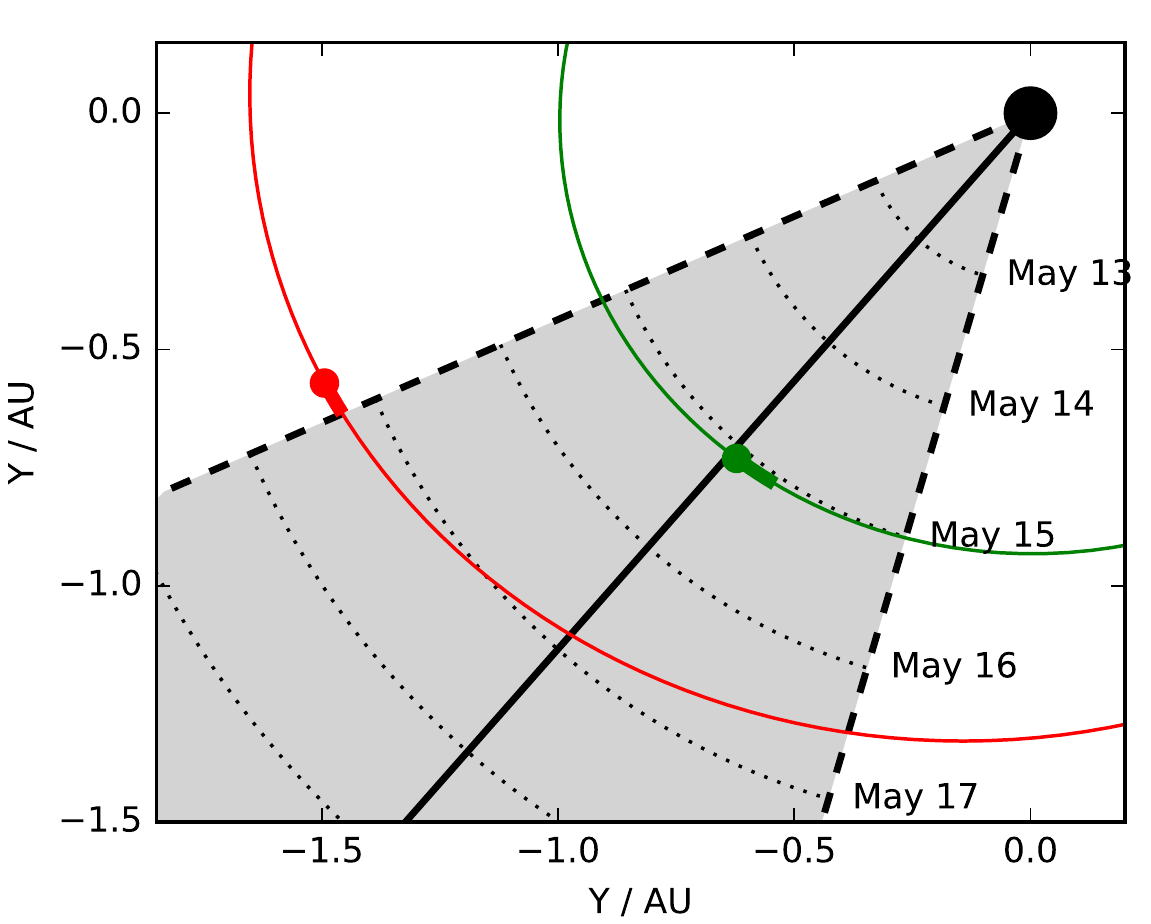}
	\caption{Position of Mars (red) and Earth (Green) between 12 and 17 May 1997, in ecliptic J2000 coordinates. Thin colored lines depict the orbit of each planet, while the thicker portions indicate their motion during this interval, from their starting positions indicated by the colored circles.
	The solid black line marks the propagation direction of the ICME launched on 12 May, and the dashed lines its expected azimuthal boundaries.}
	\label{fig:97}
\end{figure}

Observed from Earth, this was a classic example of a so-called ``halo'' ICME,
and was widely studied and modeled by several groups~\citep{arge04a, odstrcil04a, odstrcil05a, wu07a, zhou08a, cohen10a}.
It was estimated as having an angular diameter of $\sim50^\circ$, with the direction of propagation located within 1$^\circ$ of the Sun-Earth line~\citep{odstrcil04a}.
At this time, the azimuthal separation between Earth and Mars was $\sim$30$^\circ$, as depicted in Figure~\ref{fig:97} where the orbits of Earth and Mars are shown by the green and red lines, respectively, in the ecliptic J2000 coordinate system.
Colored circles indicate the position of each planet at the time the ICME was launched on 12 May 1997, while the thicker lines indicate their respective orbital motion to 17 May 1997.
The propagation direction of the ICME is shown by the black solid line, only slightly displaced from the Sun-Earth line, and the expected azimuthal extent of the ICME is indicated by the gray shaded region.
The progression of the ICME front is approximately indicated by the dotted arcs and the adjacent day numbers, based on the results of~\citep{odstrcil05a}.

On the basis of these studies we conclude that it is likely that Mars would have also experienced a significant solar wind disturbance due to the impact of the flank of this ICME, which we anticipate to have occurred within a few hours of midnight (00 UTC) on 17 May 1997, based on the $\sim$500~$km\;s^{-1}$ velocity of the ICME front measured as it reached Earth.
This would place the impact a matter of hours before the HST observations of the same day.
The MSWIM solar wind propagation was also inspected for this period, which yielded a somewhat earlier arrival time for the shock than that expected from the studies of the halo ICME at Earth by~\cite{odstrcil05a} and others.
This earlier predicted arrival time from MSWIM is consistent with the limitations of the propagation method itself, which will generally yield an earlier arrival time for an ICME-like structure for the relative positioning of Mars and Earth shown in Figure~\ref{fig:97}.
Specifically, MSWIM predicted the arrival of the shock at around ~12 h UTC on 16 May, i.e. still in advance of the HST observation and approximately a half-day earlier than depicted in Figure~\ref{fig:97}.

Measurements of the angular extent of the ICME front cannot be further constrained with available data, but we note that a shift of the propagation direction of the ICME, or increase of its azimuthal extent by only a few degrees would likely increase the magnitude of the disturbance expected at Mars.
We note that while this particular ICME propagated into a relatively undisturbed preceding solar wind, the potentially complex evolution of the magnetic fields during its early expansion has been studied in detail~\citep{cohen10a}, which may be relevant to its parameters once it emerges into the heliosphere which are not captured by the simple `cone' approximation depicted in Figure~\ref{fig:97}.
The bulk parameters of the ICME may also vary significantly along its azimuthal extent.
In conclusion, while this further tentative association of a Mars atmospheric plume with a preceding ICME impact proves nothing outright, it does lend further weight to a possible direct connection between these two phenomena.

\subsection{Comparison with other observing intervals}

\begin{figure*}[htp]
	\includegraphics[width=0.9\linewidth]{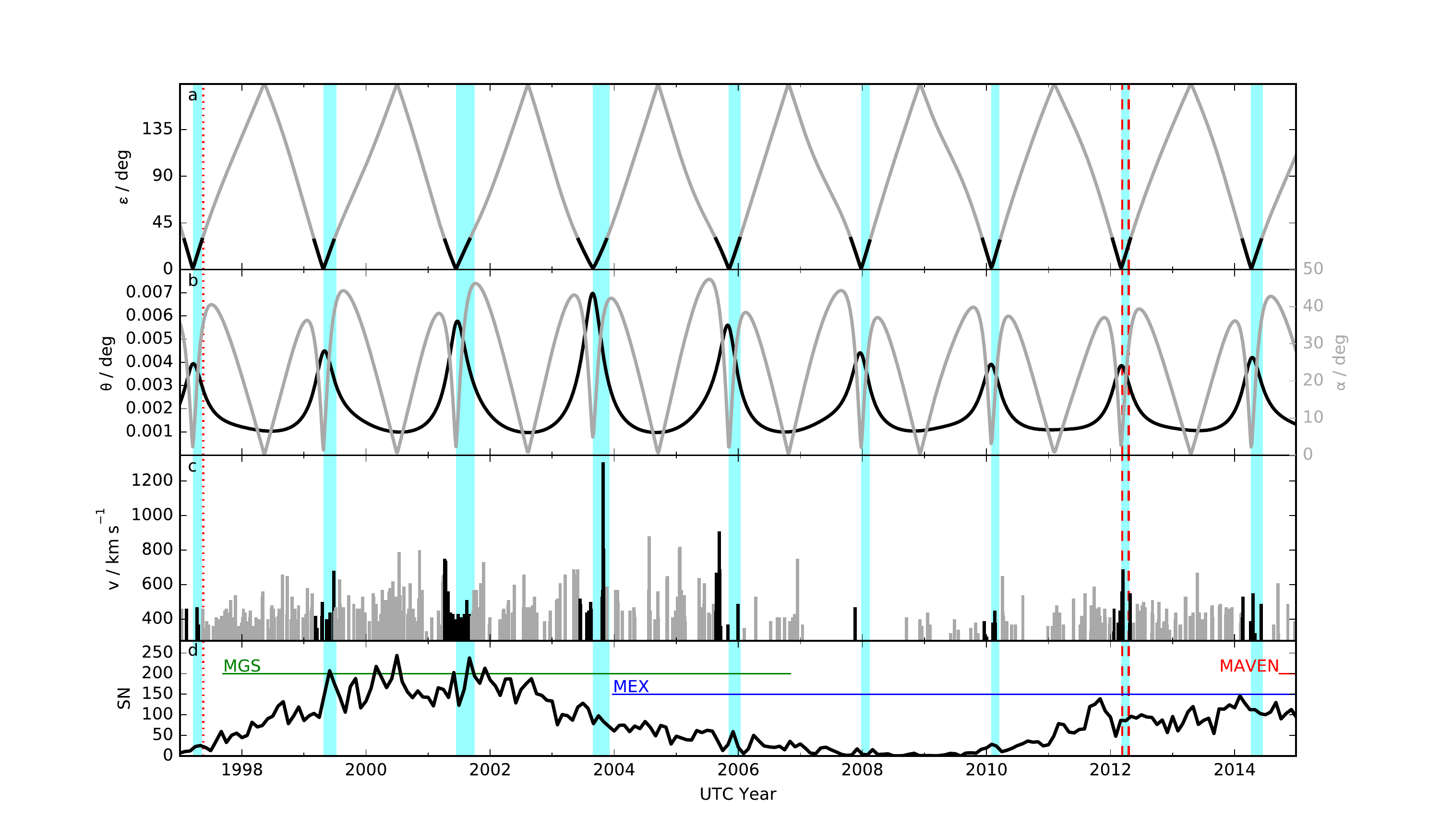}
	\caption{Comparison of timings and rates of Earth-impacting ICMEs and Mars viewing geometry.
	a) Angular separation $\epsilon$ between Mars and Earth (grey line).  Black segments indicate those intervals with $\epsilon<30^\circ$.
	b) Angular diameter $\theta$ of Mars as seen from Earth (black line, left axis), and Sun - Mars - Earth phase angle $\alpha$ (grey line, right axis).
	c) Average velocity measured in ICMEs observed at Earth, as given by \cite{cane03a} and \citep{richardson10a} (grey lines).  Individual ICMEs are highlighted black for those events occurring during periods with $\epsilon<30^\circ$.
	d) Monthly-averaged sun spot number SN, recorded by the Royal Observatory of Belgium. Overplotted horizontal colored bars indicate the durations of scientific measurements made by the labelled missions at Mars.
	Vertical red dashed lines bound the period for which the plume was observed by \emph{SL15}.
	Vertical shaded blue regions indicate all periods for which the viewing conditions of Mars were identical, or better, than in this interval.
	The vertical red dotted line marks the timing of the 1997 plume observation made by Hubble, as discussed by \emph{SL15}.}
	\label{fig:cmes}
\end{figure*}

The obvious question remains - if these plumes are in some way the result of the impact of large ICMEs upon the Martian system, why have they not been observed more frequently?
In Figure~\ref{fig:cmes} we compare a catalogue of ICMEs observed at Earth, the viewing geometry of Mars, and the progression of the solar cycle throughout this century.
The grey trace in panel a of Figure~\ref{fig:cmes} shows the angle $\epsilon$ between Mars and Earth in the heliosphere, with $\epsilon = 0^\circ$ when the planets are radially aligned.
The black lines highlight those periods for which $\epsilon < 30^\circ$, i.e. a condition similar to or better than the configuation during the HST observations discussed above.
Panel b then show the angular diameter $\theta$ of Mars as viewed from Earth and the phase angle $\alpha$ (the Sun--Mars--Earth angle, shown grey and referenced to the right axis).
Blue shaded regions throughout the figure then indicate periods with quantitively similar viewing conditions to those afforded during the plume observations reported by \emph{SL15}.
Specifically, these are periods with both $\theta>0.003^\circ$ and $\alpha<30^\circ$ and increasing with time, corresponding to visibility of the dawn terminator from Earth.

In panel c of Figure~\ref{fig:cmes} we plot the average velocity of ICMEs in the list published by~\citep{cane03a, richardson10a}\footnote{Obtained via \href{http://www.srl.caltech.edu/ACE/ASC/DATA/level3/icmetable2.htm}{the ACE Science Center}.}, detected at Earth by various spacecraft.
Each recorded CME is shown by a vertical line, colored black for those intervals when Mars and Earth are closely aligned with $\epsilon < 30^\circ$.
Finally, panel d of Figure~\ref{fig:cmes} shows the monthly averaged sun spot number.
For reference, the colored and labelled lines in this panel show the intervals spanned by the MEX, MGS and MAVEN missions.
In each panel, the red dashed lines indicates the first and last detections of the plume reported by \emph{SL15}, while the red dotted line indicates the timing of the plume observation made using HST.

The rate of occurrence of fast ICMEs varies with the solar cycle.
While that which arrived at Mars in March 2012 was one of the most significant events to occur during the MEX mission, is was also far from unique in its intensity.
For example, events with similar speeds occurred frequently during the first years of the MEX mission, 2004-2005, along with a single similar event during 2006.
The vast majority of these recorded events at Earth are not expected to impact Mars, as tentatively indicated by their grey colors in panel~c of Figure~\ref{fig:cmes}.
However, those observed at Earth around apparent opposition are significantly more likely to impact Mars, i.e. periods for which $\epsilon$ is small and Mars appears larger in the sky as shown in panels a and b of Figure~\ref{fig:cmes}.
Outside of these intervals, even very azimuthally extended ICMEs seen at Earth will likely not impact Mars.
We show this particular catalogue of Earth-impacting ICMEs here because it is derived from continuous, dedicated solar wind measurements with a consistent data set over more than a decade, something unfortunately not possible with current solar wind measurements made intermittently at Mars.

The ICME that struck Mars immediately before the plume observations in 2012 was the strongest to have impacted Mars under similar viewing conditions from Earth, apart from the extreme ICME associated with the 2003 ``Halloween storm'' at Earth.
The effects of the Halloween storm event upon the Martian plasma environment were studied by~\cite{crider05a}, with the extreme compression of the induced magnetosphere clearly visible in measurements made by MGS.
However, MGS lacked much of the plasma instrumentation available on MEX, and in particular had no ability to determine ion plasma density and composition, nor the state of the sub-spacecraft ionosphere.
While Mars was somewhat larger in the sky at this stage in 2003, the phase angle was significantly larger than for the event in 2012 ($\sim$35$^\circ$ versus $\sim$5$^\circ$ at onset), which may significantly alter the visibility from Earth of any features at high altitudes beyond the terminator.
No plume was observed associated with this extreme event, either by orbiting spacecraft or in the set of $\sim$3500 amateur optical images surveyed by \emph{SL15} obtained during the observing seasons of 2001 - 2014.
It must be noted, however, that both the quality and quantity of such amateur observations are much improved in more recent observations.
The lack of an observed plume for this event may simply reflect this fact.
We also note that the Earth-impacting Halo ICME we suggest may be related to the HST observed plume event in 1997 is, at least in terms of its average velocity, not an extreme event in comparison to other ICMEs present in this catalogue.
This may also suggest that the average velocity may not be a controlling factor in the formation of a plume.
Additionally, both this and the plume of 2012 were detected during northern summer conditions on Mars, while the more extreme ICME of 2003 impacted Mars during northern winter, and thus the typical plasma conditions over the intense southern hemisphere crustal fields may well be significantly different, leading to a potentially different response.

Similarly, an ICME of moderate intensity may have been expected at Mars on $\sim$21-22 April 2014, with average velocity of $\sim$500 km$\;$s$^{-1}$, yet no plume was reported.
For the majority of the MEX mission at Mars, the intervals with similar viewing conditions to those of March - April 2012 have been marked by the absence of ICMEs entirely (e.g., 2008), or by only relative weak events (2005/6, 2010, 2014).
\emph{SL15} do report ``occasional'' observations of near-terminator clouds seen at the limb, at altitudes that are more comparable to those seen by dedicated in-orbit observations by spacecraft.
However, specific times of such observations are not given, and there may be no relationship between these lower-altitude clouds and the extreme altitude occurrences studied here.

The potential significance of these atmospheric plumes remains to be quantified.
Any process acting to loft large amounts of material to altitudes where it is more able to escape the atmosphere, in response to extreme solar wind driving could potentially be a major contributor to the evolution of the planet's atmosphere.
The typically elevated plasma densities seen in these regions of crustal fields, at all altitudes studied may be further influenced by the passage of a ICME.
For example, the ionospheric upwellings studied by~\cite{gurnett05a} and~\cite{duru06a}, while having been shown to be stable features of the ionosphere~\citep{andrews14a}, may be enhanced during extreme events due to increased ionospheric Joule heating.
However, quite how such heating and elevation of the ionosphere may lead to such a significant vertical transport of relatively massive dust or ice particles from much lower altitudes remains to be investigated.
Similarly, electrostatic forces may become significant in this region, but whether they can ever reach sufficient strength to strongly influence the dynamics of water ice grains remains to be studied.
While micro-meter sized grains posited by \emph{SL15} as one possible explanation for the optical observations will become negatively charged in the ionosphere at these altitudes, electrostatic fields many orders of magnitude larger than those typically expected in the Martian ionosphere would still be required to balance these grains against gravity.
Strong electric fields may be present in localised regions of the Martian ionosphere, as a result of steep gradients in the ionospheric conductivity around regions of intense crustal fields.
These fields may be particularly strong during and following the impact of a fast ICME, as ionospheric plasma flows may be significantly enhanced in response to the disturbance, and consequently act to loft a localized dust-loaded region of the atmosphere to the observed high altitudes.

Finally, we note that NASA's MAVEN mission is now sampling this range of altitudes in-situ, with a comprehensive suite of science instruments, and should hopefully be able to make more conclusive statements about this phenomenon should it occur again.
In particular, valuable information can be gained from in-situ measurements of magnetic field gradients, bulk parameters of the thermal plasma, and even potentially dust particle impacts recorded by the Langmuir probe antennas~\citep{andersson15b}.
MEX remains in excellent health, and data from the instruments studied here continues to be taken.
Recently, the catalogue of apoapsis images obtained with the Visual Monitoring Camera onboard MEX was publicly released, and efforts are on-going to search this new data set for similar plume observations.

\section{Conclusions}
We now briefly recap only those relatively firm conclusions drawn from our analyses of this interesting event.
\begin{enumerate}
	\item Multiple, independent observations of the Mars atmospheric plume were made by \emph{SL15} at the dawn terminator, along with several nondetections. However, the lack of continuous observations of Mars prevents conclusive statements of the timing of the start, end and duration of this plume.
	\item The ground-based observations are consistent with a) the continuous presence of a plume with its visibility controlled purely by geometric factors (only a single observation made in the second series in April is inconsistent with this conclusion), and b) a time-variable plume intermittently appearing and disappearing.
	\item No signatures are seen in the MARSIS ionospheric sounding observations over the plume location as it traverses the dusk terminator which can be firmly associated with the presence of a plume (anomalous, or unusual features in these ionograms can be understood as being purely due to the presence of strong crustal fields at this location, as extensively studied previously).
	\item Observations of highly accelerated planetary ions in ASPERA-3 data at the same location on several orbits are themselves unusual, and require further investigation and explanation.
	\item For all the observations (except one) reported by \emph{SL15}, including the plume observed by HST in May 1997, a significant ICME can be shown to have impacted Mars in the preceding days.
	\item All observations reported by \emph{SL15}, with the exception of the 1997 event, were shown to have occurred over a region of intense crustal magnetic fields, although the nature of the observations does not allow us to be more precise about the exact magnetic topology (closed arcades or open cusps).
	\item With the notable exception of the large ICME that was shown to impact Mars during the 2003 observation season, the lack of other plume detections since 2000 could be the result of the general absence of other ICME impacts during these periods of favorable viewing geometry.
	\item If, with the caveats above, these plumes were in fact driven by space-weather disturbances at Mars, this would be a truly unique discovery, without physical explanation, and potentially of great significance in the debate regarding the loss to space of the Martian atmosphere.
	\item A significant argument now exists for future monitoring of the Martian atmosphere during extreme space weather events, using available remote Earth-based observations along side in-situ optical and plasma measurements.
\end{enumerate}

\section*{Acknowledgments}
Work at IRF was supported by grants from the Swedish National Space Board (DNR 162/14) and the Swedish Research Council (DNR 621-2014-5526)
Work at Iowa was supported by NASA through contract 1224107 from the Jet Propulsion Laboratory.  Work at Leicester was supported by STFC grants ST/K001000/1 and ST/K502121/1.
MJW: the results reported herein benefitted from participation in NASA's Nexus for Exoplanet System Science (NExSS) research coordination network sponsored
by NASA's Science Mission Directorate.
We thank A. Sanchez-Lavega for the provision of summaries of raw data.
We thank K.C. Hansen and B. Zieger for providing solar wind propagations from their \href{http://mswim.engin.umich.edu/}{Michigan Solar Wind Model}.
Finally, we thank both referees for their helpful comments.

\bibliographystyle{agufull08}

\newpage

\end{document}